\documentclass[letter]{aa} % for the letters 
\newcommand{\grp}     {${\rlap.}^{\circ}$}

\newcommand{\ltsima} {$\; \buildrel < \over \sim \;$}
\newcommand{\simlt}  {\lower.5ex\hbox{\ltsima}}            % < over MMM
\newcommand{\gtsima} {$\; \buildrel > \over \sim \;$}
\newcommand{\simgt}  {\lower.5ex\hbox{\gtsima}}            % > over MMM

\newcommand{\hess}    {HESS J1848$-$018}
\newcommand{\pri}   {${\rlap.}^{\prime \prime}$}
\newcommand{\prs}   {${\rlap.}^{s}$}

\usepackage{graphicx}
\usepackage{txfonts}
\begin{document}
   \title{Infrared and radio study of the W43 cluster:}
   \subtitle{resolved binaries and non-thermal emission}
\author{
P.~L. Luque-Escamilla\inst{1,3}
A.~J. Mu\~noz-Arjonilla\inst{2,3}
\and J.~R. S\'anchez-Sutil\inst{3}
\and J. Mart\'{\i}\inst{2,3}
\and J.~A. Combi\inst{4}
\and E. S\'anchez-Ayaso\inst{2,3}}

%\offprints{P.~L. Luque-Escamilla}

\institute{
Dpto. de Ing. Mec\'anica y Minera, EPSJ,
Universidad de Ja\'en, Campus Las Lagunillas s/n, Edif. A3, 23071 Ja\'en, Spain \\
\email{peter@ujaen.es}
\and
Departamento de F\'{\i}sica, EPSJ,
Universidad de Ja\'en, Campus Las Lagunillas s/n, Edif. A3, 23071 Ja\'en, Spain \\
\email{ajmunoz@ujaen.es, jmarti@ujaen.es, esayaso@ujaen.es}
\and
Grupo de Investigaci\'on FQM-322,
Universidad de Ja\'en, Campus Las Lagunillas s/n, Edif. A3, 23071 Ja\'en, Spain \\
\email{jrssutil@ujaen.es}
\and
Instituto Argentino de Radioastronom\'{\i}a (CCT La Plata, CONICET), 
C.C.5, (1894) Villa Elisa, Buenos Aires, Argentina \\
\email{jcombi@fcaglp.unlp.edu.ar}
}

\date{Received / Accepted}

% \abstract{}{}{}{}{} 
% 5 {} token are mandatory
 
 \abstract
% Context
{The recent detection of very high-energy (VHE) gamma-ray emission from the direction of the W43 star-forming region prompted
us to investigate its stellar population in detail in an attempt to see wether or not it is possible an association.}
% Aims
{We search for the possible counterpart(s) of the gamma-ray source or any hints of them,
such as non-thermal synchrotron emission as a tracer of relativistic particles often
involved in plausible physical scenarios for VHE emission.}
%Methods
{We data-mined several archival databases with different degrees of success. The most significant results came from radio and near-infrared archival data.}
%Results
{The previously known Wolf-Rayet star in the W43 central cluster and another cluster member appear to be
resolved into two components,suggesting a likely binary nature. In addition,  extended radio emission with a 
clearly negative spectral index is detected in coincidence with the W43 cluster.  These findings could have
important implications for possible gamma-ray emitting scenarios, which we also briefly discuss.}
{}


 \titlerunning{IR and radio study of the W43 cluster}
 
      \maketitle
%
%________________________________________________________________

\section{Introduction}

W43 (=G30.8$-$0.2)  is a star-forming complex first detected at radio wavelengths by \cite{w58}. Recently, 
attention has been focused on it, because it is believed to be one of the closest starburst regions in our galaxy
(see \cite{b2010} and references therein). 
The distance to W43 was originally estimated based on H\ion{I}~absorption and reddening measurements
(\cite{sbm1978, blum99}) pointing to a 5.5 kpc value, although with a significant uncertainty. An improved estimate using 6.7 GHz
methanol masers yielded a more reliable average systemic velocity of 97.5 km $s^{-1}$, corresponding to a larger distance 
 of $9.0 \pm 0.4$ kpc (\cite{pmg2008}).  A recent revision of survey molecular emission from W43 confirms a distance consistent with the 6 to 9 kpc
 range, with preference toward the lower limit, and a total mass of about $7\times 10^6$ $M_{\odot}$ (\cite{nguyen}).
 We also refer the reader to this work for a wide-field view of the W43 as part of a very extended radio source in the galactic plane.
 From the cluster color-color diagram, \cite{blum99} derived a considerable extinction value of
 $A_V \simeq 34$ mag, equivalent to a hydrogen column density of about $6.5 \times 10^{22}$ cm$^{-2}$.

\begin{figure*}
   \centering
    \includegraphics[angle=0,width=15.0cm]{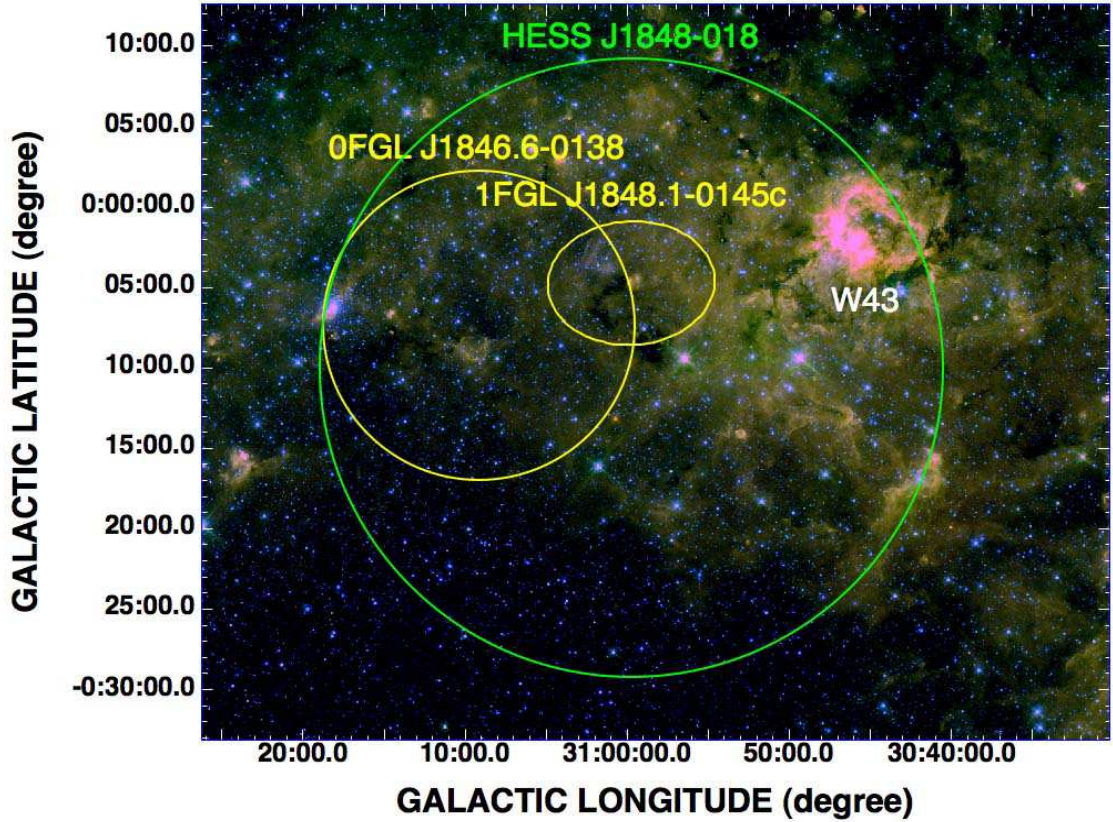}
     \caption{Composite trichromatic image of the W43 cluster the GLIMPSE 3.6, 5.8 and 8 $\mu$m bands in blue, green and red color, respectively. 
     The green and yellow ellipses indicate the positions of the H.E.S.S. and Fermi LAT sources in the field. The W43 star-forming region,
     consistent with the H.E.S.S. position, is also indicated with a white label.}
                \label{tricromia+glimpse}
    \end{figure*}

The GLIMPSE image displayed in Fig. \ref{tricromia+glimpse} 
%is first shown here to provide the reader with a general introductory view of 
shows the W43 field in the infrared band.
Among other components, W43 contains a central open cluster of massive stars illuminating
a giant H\ion{II}~region, which will be better displayed below.
A decade ago, \cite{blum99}  studied the W43 cluster in detail showing that its
three brightest members include one Wolf-Rayet (WR) and two luminous O-type stars. 
These objects are currently designated as  W43 \#1, W43 \#2 and W43 \#3, respectively. 
The same authors classified the first of them as of WN7+abs spectral type and suggested that the presence of
absorption features could be related to an unresolved companion.
W43 \#1 was subsequently  included in the 7th Catalogue of Galactic Wolf-Rayet stars (\cite{vdh01}),
 where it is also listed as WR 121a.

%\bicho\ is an obscured Wolf-Rayet (WR) star originally discovered by  \cite{blum99} and later
%included in the Seventh Catalogue of Galactic Wolf-Rayet stars (\cite{vdh01}). It is located in the
%Galaxy plane near galactic longitude $31^{\circ}$, in the central open cluster
%of a luminous star forming complex known as W43 (=G30.8$-$0.2)
%first detected at radio wavelengths by \cite{w58}.

   \begin{figure*}
   \centering
    \includegraphics[angle=0,width=15.0cm]{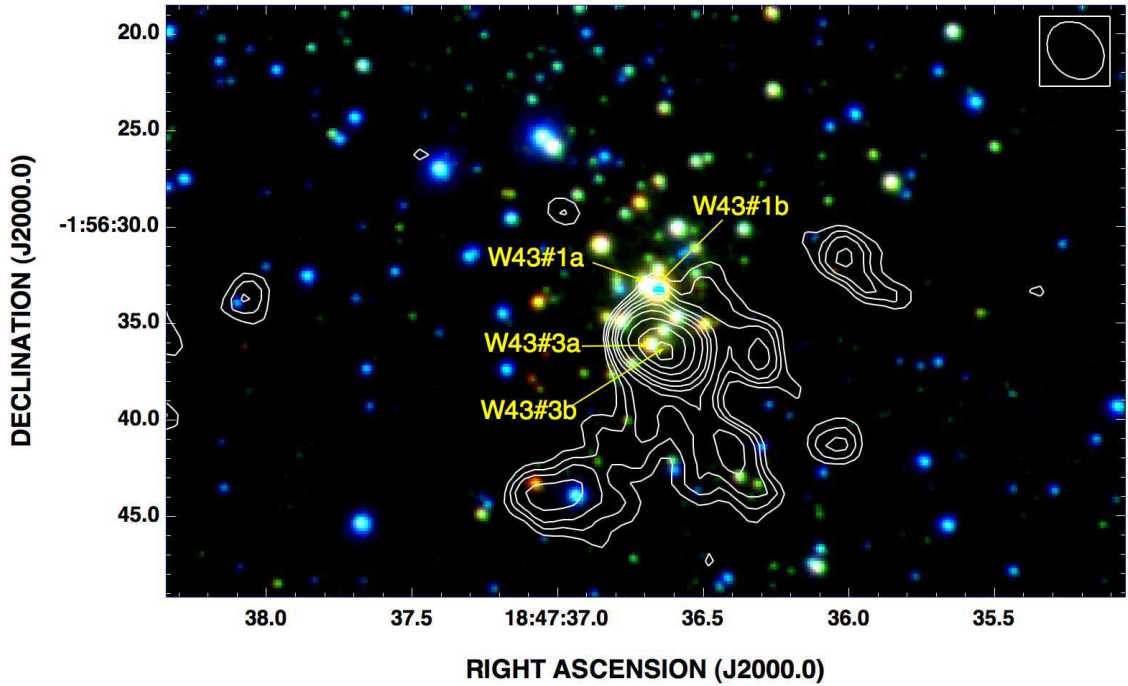}
     \caption{Composite trichromatic image of the W43 cluster
     with the ESO VLT $J$, $H$ and $Ks$ bands in blue, green and red color, respectively. The positions of the
     W43 \#1 and W43 \#3 stars proposed binary components are indicated by arrows. White contours correspond
     to 6 cm radio emission as observed with the VLA in its extended B-configuration. Contours shown correspond to
      4, 5, 6, 8, 10, 12, 15, 20, 25 and 30 times the rms noise of 0.76 mJy beam $^{-1}$. The VLA synthesized beam
      is plotted as an ellipse at the top right corner and corresponds
      to 1\pri 63 $\times$ 1\pri 31, with position angle of 42\grp 45.}
                \label{tricromia+radio}
    \end{figure*}

The W43 cluster is remarkably coincident in position with a VHE gamma-ray source 
detected by the H.E.S.S. collaboration at TeV energies (\cite{chav2009}). This VHE object is known as \hess\ and it appears
to be clearly extended with respect to the point-spread function of this Imaging Atmospheric Cherenkov Telescope array.
If the association with the cluster is correct, this will be the second proposed case of gamma-ray emission
from a stellar cluster after the well known Westerlund 2 (\cite{aha2007}). W43 \#1 (=WR 121a) has been also considered
by the \hess\ discoverers as a possible counterpart candidate. Another gamma-ray source detected by the Fermi Large Area Telescope (LAT)
is also present in the field (\cite{0FGL,1FGL}).  However, its latest available LAT error box 
corresponding to the 1FGL J1848.1$-$0145c entry in the Fermi catalog excludes 
the W43 position at present.
Whether the H.E.S.S. source is connected with the Fermi GeV emission or with the W43 stellar cluster 
still remains to be confirmed. Fermi has a strong background uncertainty in this region, which could imply
significant position shifts in future improved versions of its catalog.

In this context, we report here new observational results about
W43 mostly based on archival data. First, we present near-infrared observations
that provide  strong evidence for the binary nature of the cluster WR star and another likely member as well.
Secondly, we present the detection of extended non-thermal radio emission apparently coming from 
the cluster itself. Both could have strong implications
in our understanding of possible gamma-ray emitting scenarios, and the third part
of this paper is devoted to discuss this briefly.

\section{Analysis of ESO archive infrared observations}

We retrieved different observing epochs of W43 from the European Southern Observatory (ESO)
archives. The position of W43 was covered several times with the UT1 unit of  the ESO Very Large Telescope (VLT)
and its ISAAC instrument. In particular, we downloaded archive images obtained the on 2001 July 10 in the $J$- and $Ks$-bands under especially
good seeing conditions (0\pri 4). Data reduction was performed by means of the IRAF package mainly including flat-field, sky-background subtraction and median combining of individual frames. An astrometric solution was finally computed for the final images
using between 20 and 30 unsaturated stars. Their reference positions were taken from the Two Micron All Sky Survey (2MASS, \cite{2mass}).
The residuals of the astrometric fit had an rms value of $\pm$0\pri 09 and $\pm$0\pri 10 in right ascension and declination, respectively.
In Fig. \ref{tricromia+radio} we show the resulting composite-color image for the W43 cluster together with radio contours that will be discussed below.
The angular resolution of this  VLT infrared image fairly exceeds those previously published thanks to the excellent quality of the seeing.
This circumstance allows us to clearly distinguish the binary nature of the WR star W43 \#1 as seen in the enlarged $J$-band view shown in 
Fig. \ref{w43_1}. Hereafter, the individual components of this system will be referred to as W43 \#1a and W43 \#1b. They are separated
by $598\pm3$ mas with position angle of $255^{\circ} \pm 1^{\circ}$.
 In addition, we also detected another likely faint companion to star W43 \#3 (see Fig. \ref{w43_3} with proposed components labeled  W43 \#3a and W43 \#3b).
Their angular separation is $640 \pm 100$ mas with position angle of $271^{\circ} \pm 7^{\circ}$.

The 2MASS stars were also used to provide a zero-point photometric calibration and derive the magnitudes listed in Table \ref{posmag}
for the apparent binaries quoted above together with their positions.
Typical errors here are $\pm$0\pri 1 for each coordinate and $\pm0.1$ mag for the photometry, which was especially difficult because of the
proximity of each star pair. Magnitudes not listed in Table \ref{posmag} correspond to saturated or undetected stars.

%_____________________________________________________________
%                                             Simple A&A Table
%_____________________________________________________________
%
\begin{table} 
\caption{Observed near-infrared positions and magnitudes}             % title of Table
\label{posmag}      % is used to refer this table in the text
\centering                          % used for centering table
\begin{tabular}{cccccc}        % centered columns (4 columns)
\hline\hline                 % inserts double horizontal lines
Object &   $\alpha_{\rm J2000.0}$ &  $\delta_{J2000.0}$ &   $J$    &   $H$     &   $Ks$  \\    % table heading 
\hline                        % inserts single horizontal line
W43 \#1a    & $18^h 47^m 36$\prs 691 & $-01^{\circ} 56^{\prime}  33$\pri 06 & 16.1     &    $-$      & $-$  \\
W43 \#1b    & $18^h 47^m 36$\prs 653 & $-01^{\circ} 56^{\prime}  33$\pri 22  & 15.8    &     $-$     & $-$   \\
\hline
W43 \# 3a   & $18^h 47^m 36$\prs 677 & $-01^{\circ} 56^{\prime}  36$\pri 11  &  17.6   &  13.5      &  11.5 \\                          
W43 \# 3b   & $18^h 47^m 36$\prs 648 & $-01^{\circ} 56^{\prime}  36$\pri 08  &    $-$   &   $-$       &   13.6 \\
\hline                                   %inserts single line
\end{tabular}
\end{table}

\section{Analysis of the NRAO archive radio observations}

\begin{figure}
   \centering
    \includegraphics[angle=0,width=9.0cm]{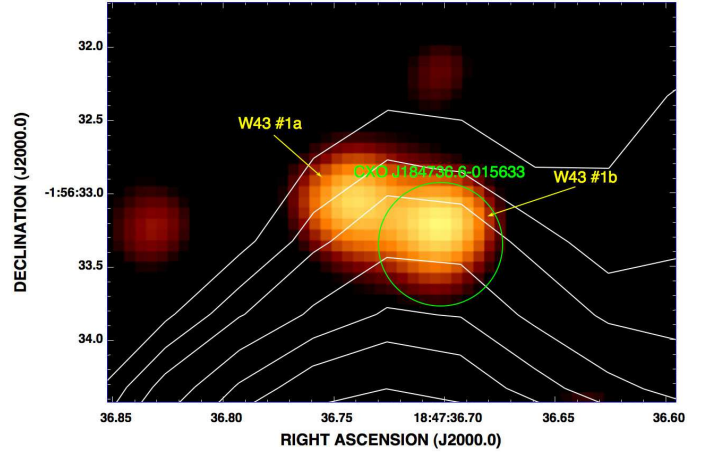}
     \caption{Zoomed $J$-band view of the WR star W43 \#1 with its components a and b clearly resolved. The green circle represents the CXO J184736.6-015633  position at the 95\% confidence level according to the Chandra X-ray source catalog.
 Radio contours are the same as in Fig. \ref{tricromia+radio}.}
              \label{w43_1}
    \end{figure}

\begin{figure}
   \centering
    \includegraphics[angle=0,width=9.0cm]{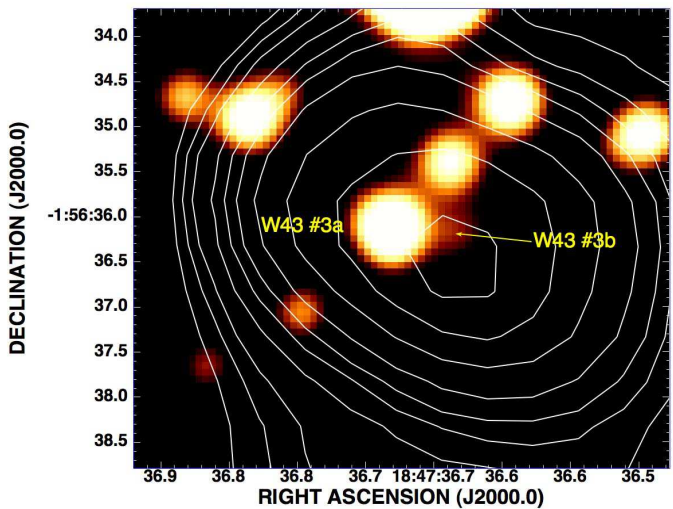}
     \caption{Zoomed $Ks$-band view of the O-type star W43 \#3 with its proposed faint companion shown by
     an arrow. Radio contours are the same as  in Fig. \ref{tricromia+radio}. }
              \label{w43_3}
    \end{figure}

Several sets of radio observations obtained with the Very Large Array (VLA) of the
National Radio Astronomy Observatory (NRAO) in the USA were used in this work.
They were either retrieved from public surveys such as MAGPIS (\cite{magpis}),
or directly downloaded and calibrated from the NRAO public archives. The AIPS
software package and standard interferometer calibration techniques were used for this purpose.
Table \ref{w43radio} lists the details of VLA observations
together with the flux densities measured for the radio source detected in the direction
of the W43 central cluster. 
This object has a clear extended morphology at all wavelengths, and we include a contour plot of our best
high angular resolution map at the 6 cm wavelength in Fig. \ref{tricromia+radio}. 
Part of these contours also appear in Figs. \ref{w43_1} and \ref{w43_3}. The central condensation
of the radio emission has a deconvolved angular size of (3\pri 8$\pm$0\pri 1)$\times$(3\pri 1$\pm$0\pri 1), with position
angle of  $61^{\circ}\pm8^{\circ}$. The J2000.0 position of the radio peak is centered at
$18^h 47^m 36$\prs 63$\pm$0\prs 01 and $-01^{\circ} 56^{\prime}$36\pri 3$\pm$0\pri 1.

%_____________________________________________________________
%                                             Simple A&A Table
%_____________________________________________________________
%
\begin{table} 
\caption{Radio emission from the W43 central cluster}             % title of Table
\label{w43radio}      % is used to refer this table in the text
\centering                          % used for centering table
\begin{tabular}{lcccc}        % centered columns (4 columns)
\hline\hline                 % inserts double horizontal lines
Data  &   Observation &  VLA                   & Frequency &  Flux Density   \\    % table heading 
 origin &    epoch          & Conf.   &   (GHz)        &      (Jy)         \\
\hline                        % inserts single horizontal line
MAGPIS     &  2007 Apr 25 & B      & 1.465  &  $0.40 \pm 0.03$ \\
AD129        & 1984 May 03  &  C    &   4.860 &  $0.25 \pm 0.03$ \\
                   & 1985 May 17  &  B    & 4.866   &         $-$             \\
AM663       & 2000 Jul 26  &   D    &   8.460 & $0.17 \pm 0.02$ \\
\hline                                   %inserts single line
\end{tabular}
\end{table}

  \begin{figure}
   \centering
    \includegraphics[angle=0,width=9.0cm]{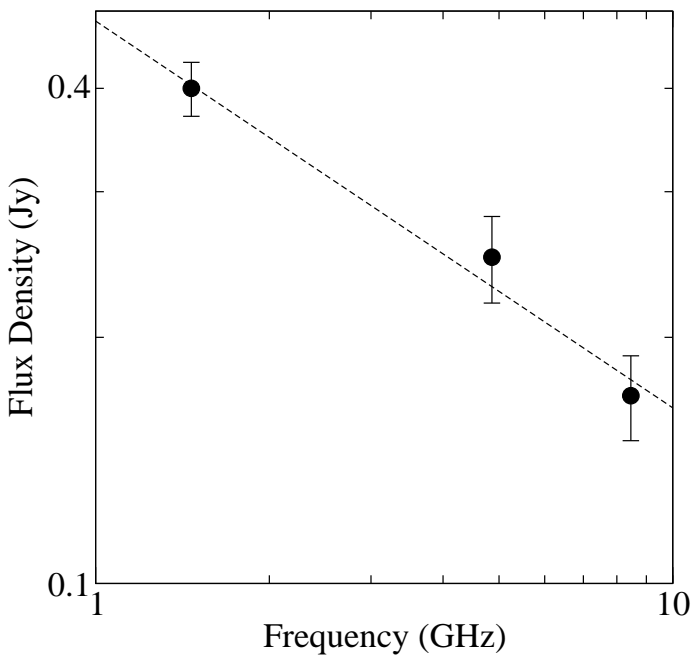}
     \caption{Radio spectrum of the W43 central cluster based on different
     sets of VLA observations as explained in the text. The strainght line is a simple
     power-law fit to the Table \ref{w43radio} flux
     densities that clearly yields a negative spectral index. \label{radio_spectrum} }
    \end{figure}

Besides obtaining a high angular resolution map, data sets in Table\ref{w43radio}
were selected  to ensure nearly matching-beam observations
at three different wavelengths. Having similar synthesized beams at all wavelengths is the best approach
to measure reliable spectral indices for any object in the field.  The radio source coincident with the W43 cluster appears
to be clearly non-thermal, as  shown in the Fig. \ref{radio_spectrum} spectrum. A simple power-law fit indicates that the observed radio emission
is well represented by $S_{\nu} = (0.48 \pm 0.02~{\rm Jy}) \left[\nu/{\rm GHz}\right]^{-0.47\pm 0.04}$. The total radio luminosity integrated from 0.1 to 100 GHz
amounts to $1.0 \times 10^{33}$ erg s$^{-1}$ with a brightness temperature as high as $7.1 \times 10^4$ K.

\section{Discussion}

The fact that the WR star W43 \#1 is now a clearly resolved binary confirms the early spectroscopic suspicion of a companion star (\cite{blum99}), 
as well as the colliding wind binary scenario recently invoked to interpret the system's X-ray emission (\cite{and2011}).
The observed close sub-arcsecond angular separation (about 0\pri 6) is comparable to other WR binaries resolved by optical telescopes
such as the Hubble Space Telescope (e.g.  \cite{niem98}). 
We have estimated the probability of a chance alignment  to be as low as $\sim5$\%, based on the 
average surface star density of the cluster and assuming a Poisson distribution.
The expected orbital period is likely to be very long, possibly decades or longer.
Indeed,  at the assumed W43 large distance the linear separation between components is likely in the range of thousands of astronomical units.
The unfortunate fact that  colors for the W43 \#1 components cannot be derived from our data owing to saturation problems in the
$H$- and $Ks$-bands prevents us from any photometric classification. However, W43 \#1b is exactly coincident with the bright X-ray
source CXO J184736.6$-$015633 detected by the {\it Chandra} satellite with a 10.8$\sigma$ confidence level (see its error circle in Fig. \ref{w43_1}).
In a WR binary system, the X-ray emitting region due to colliding winds is normally close to the star with a weaker wind. Therefore, we speculate
that W43 \#1a is likely the WR member of the system, while W43  \#1b is its O-type companion. Additional dedicated observations
should confirm this point.

It is intriguing that the peak of non-thermal radio emission from the W43 direction does not coincide with the WR system but it is very close to 
the luminous star W43 \#3 instead, nearly $3^{\prime\prime}$ to the south. This object could also be a binary
system, as already mentioned in Section 2.  Based on the distance and extinction values quoted in the introduction,
the colors of the brightest W43 \#3a component derived from our photometric data in Table \ref{posmag}  are consistent with an O-type supergiant.
For the W43 \#3b component, only a $Ks$-band absolute magnitude can be derived, which turns out to agree
with a main sequence O-type star.

Given the significantly high brightness temperature and negative spectral index observed, the W43 radio emission is likely to be 
of non-thermal synchrotron nature. Winds from luminous
stellar objects are known to be capable of accelerating relativistic electrons, from which  synchrotron
emission is naturally expected even if they are single (see e.g. \cite{vl05}). However,
from the radio spectrum in Sec. 3  the monochromatic 6 cm luminosity is $2.4 \times 10^{22}$ erg s$^{-1}$ Hz$^{-1}$,
which exceeds the average non-thermal luminosity of a typical WR star  (\cite{ch1999}) by nearly three orders of magnitude.
This, together with the clearly extended appearance of the radio source, 
suggests that the collective effect of stellar winds of the WR and O stars in the cluster is the responsible for this non-thermal radio emission. 
Indeed, the possibility that open clusters may harbor
a large-scale population of particles up to TeV energies accelerated by their strongest O star winds, and even supernova remnants, 
has already been seriously considered by recent theoretical models (see e.g. \cite{ger2010}).
In this context, the detection of non-thermal radio emission and the likely binary nature for the most luminous stars in the cluster
give support to the identification of W43 with the \hess\ gamma-ray source (\cite{chav2009}).

Assuming standard energy equipartition conditions between
the relativistic electrons and the magnetic field (\cite{pa1970}), the radio source total energy content is estimated
to be $2.2 \times 10^{45}$ erg with a magnetic field of $6.8 \times 10^{-4}$ G. Future theoretical work should ascertain whether this energy
content and magnetic field are feasible within the proposed scenario. Alternatively, we cannot exclude the possibility that the detected radio
source, whether or not it is related to the high-energy source, is of extragalactic origin. In this case, the overlapping with the cluster itself would be
a mere line-of-sight coincidence, but the probability of this being true is negligible.  
Indeed,  based on extragalactic radio-source counts (see e.g. the formulation by  \cite{lf1989}),
the expected number of radio sources as bright as the
one reported here within a solid angle similar to that of the W43 central cluster is  $\sim 10^{-6}$.

%CXO J184736.6-015633     18 47 36,65    -01 56 33,37       0,42

% Equiparticin

%   Lluminositat radio   (0.1-100 GHz)     :  0.98187D+33 erg/s
 %     Tamany de la radiofont    :  0.46346D+18 cm   
  %    Temperatura de brillantor :  0.71425D+05 K    
 %     Energia Total minima      :  0.22146D+46 erg  
 %     Camp magnetic minim       :  0.67650D-03 G    

\section{Conclusions}

The main results of this paper can be summarized as follows:

\begin{enumerate}

\item Based on archival data, we have reported the best angular resolution observations of the W43 central cluster at infrared and radio wavelengths,
which shed some light on its possible association with high-energy gamma-ray emission.

\item The binary nature of  the WR star W43 \#1 has been confirmed by resolving its components
with sub-arcsecond angular separation. One of them is clearly coincident with
the  {\it Chandra} X-ray source in the field. Another member of the cluster is also proposed to be a 
binary system consisting of two O-type stars.  

\item Non-thermal extended radio emission is clearly detected from the direction of the cluster. 
We speculate that it could be the result of a collective effect of the stellar winds from the whole population
of luminous massive stars in the W43 cluster as the most conceivable scenario.
Other alternative interpretations could also be considered and more observations are required to fully solve this issue.

\end{enumerate}

\begin{acknowledgements}
{\small 
The authors acknowledge support of different aspects of this work
 by grants AYA2010-21782-C03-03 from the Spanish Government, 
Consejer\'{\i}a de Econom\'{\i}a, Innovaci\'on y Ciencia
of Junta de Andaluc\'{\i}a as research group FQM-322 and excellence fund FQM-5418, as well as FEDER funds. 
%J.A.C. is a research 
%member of the Consejo Nacional de Investigaciones Cient\'{\i}ficas y Tecnol\'ogicas (CONICET), Argentina. 
Based on observations made with the European Southern Observatory
telescopes obtained from the ESO/ST-ECF Science Archive Facility.
The NRAO is a facility of the NSF operated under cooperative agreement by Associated Universities, Inc. 
This publication makes use of data products from the Two Micron All Sky Survey, which is a joint project 
of the University of Massachusetts and the Infrared Processing and Analysis Center/California Institute 
of Technology, funded by the National Aeronautics and Space Administration and the National Science 
Foundation in the USA. 
This research has made use of the NASA/ IPAC Infrared Science Archive, which is operated by the Jet Propulsion Laboratory, California Institute of Technology, under contract with the National Aeronautics and Space Administration,
and of the Spitzer-GLIMPSE database, which is operated by the Jet Propulsion Laboratory, California Institute of Technology.
This research has made use of data obtained from the Chandra Data Archive and the Chandra Source Catalog.
Authors also acknowledge useful comments and discussion with Prof. Gustavo E. Romero (IAR).
 
%and software provided by the Chandra X-ray Center (CXC) in the application packages CIAO, ChIPS, and Sherpa.
}
\end{acknowledgements}

\end{document}